\documentclass[10pt,conference]{IEEEtran}
\IEEEoverridecommandlockouts
\usepackage[numbers]{natbib}
\usepackage{dblfloatfix}    
\usepackage{color, colortbl}
\definecolor{Gray}{gray}{0.9}
\usepackage{dblfloatfix}    
\usepackage[utf8]{inputenc} 
\usepackage{tabularx}
\usepackage{nomencl}
\usepackage{tikz}
\usepackage{xparse}
\usepackage{lipsum}
\usepackage{subfigure}
\usepackage{scalefnt}
\usepackage{adjustbox}
\usepackage{framed}
\usepackage{float}
\usepackage{xspace}
\usepackage{comment}
\usepackage{siunitx,booktabs}
\usepackage{multirow,hhline,graphicx,array}
\usepackage[most]{tcolorbox}
\usepackage{arydshln}
\usepackage[flushleft]{threeparttable}
\usepackage[normalem]{ulem}

\usepackage[strict]{changepage}
\usepackage{framed}

\definecolor{formalshade}{rgb}{0.95,0.95,1}

\newcolumntype{L}[1]{>{\raggedright\let\newline\\\arraybackslash\hspace{0pt}}m{#1}}
\newcolumntype{C}[1]{>{\centering\let\newline\\\arraybackslash\hspace{0pt}}m{#1}}
\newcolumntype{R}[1]{>{\raggedleft\let\newline\\\arraybackslash\hspace{0pt}}m{#1}}

\newcommand{\MyBox}[1]{\vspace{3mm}\noindent\framebox[\columnwidth][c]{\parbox[b]{0.95\columnwidth}{ #1 }}\vspace{3mm}}

\usepackage{xcolor}
\definecolor{mypurple}{rgb}{139,0,139}
\definecolor{mygreen}{rgb}{0,140,0}
\definecolor{orange}{RGB}{255,127,0}
\ifCLASSINFOpdf
\else
\fi

\newif\ifdraft
\drafttrue

\newcommand{\nb}[2]{
	{
		{\color{black}{
				\small\fbox{\bfseries\sffamily\scriptsize#1}
				{\sffamily\small$\triangleright~${\it\sffamily\small #2}$~\triangleleft$}
	}}}
}

\ifdraft
\newcommand\grex[1]{\nb{Gregorio}{\color{red}#1}}
\newcommand\anita[1]{\nb{Anita}{\color{blue}#1}}
\newcommand\igor[1]{\nb{Igor}{\color{purple}#1}}
\newcommand\christoph[1]{\nb{Christoph}{\color{olive}#1}}
\newcommand\marco[1]{\nb{\color{red}Marco}{\color{blue}#1}}
\newcommand\bianca[1]{\nb{Bianca}{\color{purple}#1}}
\newcommand\georg[1]{\nb{Georg}{\color{olive}#1}}

\newcommand{\fixme}[1]{{\textcolor{red}{[FIXME] #1}}\xspace}

\else
\newcommand\grex[1]{}
\newcommand\igor[1]{}
\newcommand\christoph[1]{}
\newcommand\marco[1]{}
\newcommand\bianca[1]{}
\newcommand\anita[1]{}
\newcommand\georg[1]{}

\newcommand{\fixme}[1]{}

\fi

\newcommand\clearrow{\global\let\rowmac\relax}
\clearrow

\begin{document}

\title{The Shifting Sands of Motivation: Revisiting What Drives Contributors in Open Source}


\author{\IEEEauthorblockN{Marco Gerosa,\textsuperscript{$1$} Igor Wiese,\textsuperscript{$2$}  Bianca Trinkenreich,\textsuperscript{$1$} Georg Link,\textsuperscript{$3$} Gregorio Robles,\textsuperscript{$4$} Christoph Treude,\textsuperscript{$5$} \\Igor Steinmacher,\textsuperscript{$1, 2$} Anita Sarma\textsuperscript{$6$}}

\IEEEauthorblockA{\textsuperscript{$1$}\textit{Northern Arizona University, USA}, \textsuperscript{$2$}\textit{Universidade Tecnológica Federal do Paraná, Brazil}, \textsuperscript{$3$}\textit{Bitergia, USA},
\\ \textsuperscript{$4$}\textit{Universidad Rey Juan Carlos, Spain}, \textsuperscript{$5$}\textit{University of Adelaide, Australia}, \textsuperscript{$6$}\textit{Oregon State University, USA}
\\ 
\\
marco.gerosa@nau.edu, igor@utfpr.edu.br, bt473@nau.edu, georglink@bitergia.com, grex@gsyc.urjc.es\\christoph.treude@adelaide.edu.au, igor.steinmacher@nau.edu, anita.sarma@oregonstate.edu}
}

\maketitle
\thispagestyle{plain}
\pagestyle{plain}

\begin{abstract}
Open Source Software (OSS) has changed drastically over the last decade, with OSS projects now producing a large ecosystem of popular products, involving industry participation, and providing professional career opportunities. But our field's understanding of what motivates people to contribute to OSS is still fundamentally grounded in studies from the early 2000s. With the changed landscape of OSS, it is very likely that motivations to join OSS have also evolved. Through a survey of 242 OSS contributors, we investigate shifts in motivation from three perspectives: (1) the impact of the new OSS landscape, (2) the impact of individuals' personal growth as they become part of OSS communities, and (3) the impact of differences in individuals' demographics. Our results show that some motivations related to social aspects and reputation increased in frequency and that some intrinsic and internalized motivations, such as learning and intellectual stimulation, are still highly relevant. We also found that contributing to OSS often transforms extrinsic motivations to intrinsic, and that while experienced contributors often shift toward altruism, novices often shift toward career, fun, kinship, and learning. OSS projects can leverage our results to revisit current strategies to attract and retain contributors, and researchers and tool builders can better support the design of new studies and tools to engage and support OSS development. 
\end{abstract}


\section{Introduction}

Much has changed since the early days of Open Source Software (OSS)~\cite{steinmacher2017free}, from the type of products it creates, to who participates (e.g., individuals, industry consortia, companies), to how it operates (e.g., industry funded projects, foundations, social coding platforms). OSS today enjoys a place of distinction in producing key technologies and providing learning and career opportunities. With such drastic changes to the status of OSS, along with the clearer path to personal economic gain that it now affords, it is likely that what motivates people to join OSS has evolved since the early days.

Our current understanding of what motivates individuals to contribute to OSS, however, remains largely rooted in research from the 2000s, when OSS was still in its infancy~\cite{lakhani2003hackers, harsworking}. This dated understanding of motivation can make our community and research efforts to attract, sustain, and improve diversity in OSS projects ineffectual. Therefore, it is time we revisit the fundamental question of what drives people to contribute to OSS today, giving us a two-part research question:\\
\textit{RQ1a: What motivates contributors to OSS today?}\\
\textit{RQ1b: How has motivation to contribute shifted as OSS has matured?}

Besides understanding what motivates individuals now, so we can better support them, we also aim to identify the ways in which people's motivations have shifted in response to the changing landscape, so that OSS communities can rethink their strategies to attract and retain contributors.


Shifts in motivation occur not only because of changes to the OSS landscape, but might also reflect the journey an individual makes and their growth since first joining~\cite{steinmacher2014attracting}. Currently, we lack an understanding of the differences in motivation for the early joiners compared to those who are well-entrenched in OSS. To support both the attraction of new members and the retention of existing contributors, we need to understand how the motivation changes after the members join OSS. This leads us to our next research question:\\
\textit{RQ2: How does motivation to contribute to OSS shift as OSS contributors gain tenure?}

What motivates people and shifts their motivation as they gain experience in OSS may also depend on their individual characteristics---gender, degree of experience, primary type of contribution (code, non-code), and so on. For example, research has shown women enjoy other types of contributions (such as documentation or community management) over code hacking~\cite{robles2016women}. Research has also shown that coders and non-coders follow different career pathways~\cite{trinkenreich2020pathways}. Therefore, knowing what motivates people with different individual characteristics can help us in better supporting a diverse community. This brings us to our final (two-part) research question:\\
\textit{RQ3a: How does motivation to contribute differ for diverse characteristics?\\
RQ3b: How do shifts in motivation differ for diverse characteristics?}

To answer our research questions, we conducted an online survey, revisiting the questions used to measure motivation in three seminal papers: \citet{harsworking}, \citet{lakhani2003hackers}, and \citet{ghosh2002free}. The responses from 242 OSS contributors indicate that indeed motivation has shifted as OSS has matured and that individuals' motivations evolve as they contribute to OSS. As \citet{von2012carrots} famously reported \textit{``it is not the immediate and isolated outcome that matters (the carrot), but how the individual subjectively holds outcomes and actions to be consistent over time (the journey toward the end of the rainbow).}'' Understanding what motivates people should include not only the immediate carrots, but also the larger quest of an individual as they continue to grow and reach for the pot of gold at the end of their journey. A deeper understanding of motivation and satisfaction can help OSS projects in identifying strategies to lower contributor turnover and improve productivity, as identified by literature on motivation in software engineering~\cite{beecham2008motivation,francca2011motivation,DASILVA2012216, helen2008}. We hope that our insights on the shifting state of motivation help our research community create more nuanced approaches to attract and support a diverse set of contributors to OSS.

\section{Related Work}
\label{section:relatedwork}

Motivation has been a frequently studied topic in software engineering---a systematic literature review found 92 papers on this topic published until 2006~\cite{beecham2008motivation}. In 2010, this literature review was updated to add 53 additional papers~\cite{francca2011motivation}. The literature shows that proper management of motivation and satisfaction helps software organizations achieve higher levels of productivity, avoid turnover, budget overflows, and delivery delays \cite{beecham2008motivation,francca2011motivation,DASILVA2012216}. \citet{helen2008} provide a comprehensive overview of the motivation models used in software engineering and propose a new model by leveraging previous results from the literature. More recently, \citet{HelenESEM} analyzed data from interviews with 13 professional software engineers and suggested that there has been a trend toward more socially-oriented motivators in software engineering. Other works, such as \citet{Helen}, found a variety of factors that influence work motivation and job performance in the software industry.

However, motivation in software industry settings may not necessarily apply to OSS~\cite{roberts2006understanding}. OSS contributors have a high degree of autonomy, intrinsic motivation, and self-determination~\cite{roberts2006understanding}. 

Motivation to contribute to OSS was extensively studied in the early 2000s. Many researchers were intrigued that high-quality OSS was developed volunteerly by qualified, young, motivated individuals \cite{bitzer2007intrinsic}. \citet{harsworking}, \citet{ghosh2002free}, and \citet{lakhani2003hackers} conducted broad web-based surveys to collect motivations of OSS contributors in a two-year period from 2000-2002. Other surveys from the early 2000s focused on specific communities such as Linux \cite{hertel2003motivation} and Apache \cite{hann2004developers, roberts2006understanding}. Since these initial surveys, which are still considered state-of-the-art, researchers conducted studies focused on specific communities \cite{choi2015characteristics, spaeth2015research, bosu2019understanding}. Other researchers focused on the motivation of specific contributor profiles, such as newcomers \cite{hannebauer2017relationship}, one-time code contributors \cite{lee2017understanding}, quasi-contributors \cite{steinmacher2018almost}, and students \cite{SILVA2020110487}. Researchers also investigated the relation between motivation and other constructs, like retention \cite{wu2007empirical}, task effort \cite{ke2010effects}, intention to contribute \cite{wu2011influence}, and participation level \cite{meissonierm2012toward}. Specific types of motivations were also investigated \cite{iskoujina2015knowledge, stewart2006impact}.

\citet{von2012carrots} surveyed the literature to aggregate the studies about motivation in OSS published until 2009. They identified ten motivation categories, grouped as intrinsic, internalized-extrinsic, and extrinsic. Intrinsic motivation moves the person to act for the fun or challenge entailed rather than in response to external pressures or rewards~\cite{ryan2000self}. In contrast, extrinsic motivations are based on outside incentives when people change their actions due to an external intervention~\cite{frey1997relationship}. Developers can also internalize extrinsic motivators in a way that they are perceived as self-regulating behavior rather than external impositions \cite{deci1987support,roberts2006understanding}. These internalized extrinsic motivations include reputation, reciprocity, learning, and own-use. 

Essentially, the broad research about motivation was conducted before the professionalization of open-source development, the increasing involvement of corporations, and the rise of social coding platforms. Much of the work focused on SourceForge---GitHub was launched in 2008 and became the dominant open-source hosting site around 2012 \cite{zhou2020has}. Therefore, it is time to revisit, replicate, and extend previous research on motivation to contribute to OSS. Furthermore, investigating the shift of motivation, which is core in our study, was not the focus of any previous work.


\section{Research Method}

To answer our research questions, we administered an online survey to OSS contributors. In the following subsections, we describe our approach and instrumentation. 

\subsection{Selection of previous surveys} 

To select prior surveys to help design our study, we searched for broad surveys with a high number of citations on Google Scholar, and selected the surveys conducted by \citet{lakhani2003hackers} (2028 citations) and \citet{harsworking} (1696 citations)---the search was conducted in May 2020. We also included the survey conducted by \citet{ghosh2002free}, as it was performed during the same period as the others but collected a significantly higher number of respondents (2784; compared to 684 and 81, respectively). We searched for complementary information that was not available in the manuscripts, such as full instruments, data sets, and technical reports with more details about the studies. Additionally, we wrote to the previous surveys' authors asking for additional information; however, we did not receive an answer. 




\subsection{Identification of similar questions}
\label{method:mapping_question}

As we wanted to compare our results to the previous surveys but did not want to ask the participants redundant questions, the next step of our method was identifying similar questions. We used negotiated agreement~\cite{garrison2006revisiting} to group questions (or categories in the case of \cite{harsworking}) that could be considered similar. For example, we grouped ``Peer Recognition" (Hars), ``Enhance reputation in F/OSS community" (Lakhani), and ``Get a reputation in OS/FS community" (Ghosh). Three researchers experienced in qualitative methods and OSS performed the initial grouping of the questions. They leveraged the description provided in the original papers and supplementary materials to disambiguate the meaning of the motivation factors and questions. They kept meeting and discussing until reaching an agreement. After that, four other researchers (three academic and one practitioner) validated the grouping. We ended up with 20 questions extracted from the previous surveys. 


\subsection{Definition of motivation factors}
\label{motivation_factors}

After identifying similar questions, we decided to group them into higher-level constructs to narrow down the analysis.  To the best of our knowledge, the literature review conducted by \citet{von2012carrots} is the most comprehensive investigation of motivation in OSS, since they aggregated motivation factors found in 40 primary studies published until 2009. Therefore, we employed this work as our theoretical framework to guide our analysis. In it, the authors grouped the motivation factors in ten main categories, namely, \emph{Ideology}, \emph{Altruism}, \emph{Kinship}, \emph{Fun}, \emph{Reputation}, \emph{Reciprocity}, \emph{Learning}, \emph{Own-Use}, \emph{Career}, and \emph{Pay}. To map the questions to the categories, we followed the same process described in the previous step. When grouping the items, we noticed that the items did not completely cover the \emph{Fun} category. To cover this gap, we added one question to represent this category to our set: ``I have fun writing programs.'' Therefore, our motivation questionnaire ended up with 21 items, listed in Table~\ref{tab:literature_comparison*}.

After the informed consent, we asked two open questions about motivation to contribute to OSS. The goal was to collect spontaneous answers before presenting participants with the list of motivation factors. To understand the shift in motivation (RQ2), we asked participants why they first began and then continued contributing: (i) ``What motivated you to start contributing to Free/Open Source Software (F/OSS) projects?'' and (ii) ``Why do you continue contributing to F/OSS projects?'' As \citet{von2012carrots}'s literature review considered only papers published until 2009, these open questions also helped us to evaluate to what extent the categories proposed by \citet{von2012carrots} cover current OSS contributors' motivations to contribute. 

\subsubsection{Likert-scale questions}

On a new page, we presented the 21 items from the previous step to identify the contributors' motivations. We presented each item as an option to complement the sentence ``I contribute to F/OSS because...'' The items followed a 5-point Likert-scale (from ``Strongly agree'' to ``Strongly disagree'' with a neutral option) and ``I'm not sure.'' The list of items was randomized for each respondent to avoid ordering bias. The exact question wording is provided in our supplementary material.\footnote{https://doi.org/10.5281/zenodo.4453904} We also added an attention check item (``This is a verification question, please answer Strongly Agree.'') and a question to collect other potential motivations not covered by the items (``Are there any other reasons for contributing to F/OSS that we haven't covered above?'').


\subsubsection{Demographics}
The last part of the survey comprised demographic questions: project(s) they contribute to the most; time since the first contribution; initial and current financial relationship to OSS (paid/unpaid); initial and current type of contributions; main occupation; gender identity; country of origin and residence; and age. 

All questions were optional to increase the response rate, by making respondents more comfortable~\cite{punter2003conducting}. Moreover, to encourage participation, we offered the participants a chance to enter a raffle for US\$100. To enter the raffle case, they needed to provide an email address at the end of the survey. 

After it was proofread by all researchers and tested in multiple browsers and devices, we invited seven participants to pilot the survey so we could collect feedback and measure the time to answer. No modification of the survey was necessary, and we discarded these initial answers.

\subsection{Recruitment}
\label{section:method:dist}
Similar to previous surveys, we opted for a broad distribution of our survey. We avoided scraping email addresses from software repositories because this practice has been condemned by OSS communities~\cite{baltes2016worse} and can violate the terms of service of the platforms and some regional data protection laws. Instead, we focused on increasing the number of responses and the sample's diversity by employing several strategies. 
First, we formed an international and diverse team of researchers, who are originally from South America (4), Europe (3), and Asia (1) and were working, at the time of this study, in North America (5), Europe (1), South America (1), and Australia (1). Seven researchers work in academia with extensive experience with OSS, and one researcher is a practitioner working in an OSS company. The researchers sent direct messages to their contacts and posted ads on social network websites.


Second, we advertised the survey on social media sites, namely, Twitter, Facebook, Reddit, LinkedIn, and Hackernews. To reach a broader audience, we paid to promote our posts on Twitter, Facebook, and Reddit. These sites are largely used by our target population~\cite{singer2014software, storey2010impact, aniche2018modern}. When the site allowed, we also shared our posts to groups related to OSS development. Finally, we asked our personal contacts to share our posts and distribute our message (for example, on Twitter, our posts were retweeted more than 200 times). 

The survey was available between June 4 and July 24, 2020. We received 247 non-blank answers and, after filtering the data (as detailed in the next subsection), we ended up with 242 valid responses. 
The link posted on Twitter was the origin of almost half (46\%) of our answers. The link that we sent to our contacts (which they probably forwarded to their colleagues as we requested) was the origin of 19\% of the answers. Links posted on OSS-related discussion lists (14\%) and Reddit (10\%) were also a common origin of our responses. 


\subsection{Filtering}
\label{section:method:filtering}
We filtered our data to consider only valid responses. We dropped answers that failed the attention question (4 cases) and checked for answers with the same choice for all Likert scale questions (0 removed). We then analyzed the time to complete the survey to remove lower outliers (0 removed). We manually inspected the open text questions, looking for senseless and inappropriate answers (1 removed). Then, we filtered our data looking for potential duplicate participation, even though the survey platform (Qualtrics) has mechanisms to prevent multiples responses from the same participant. We started looking for identical and similar emails (0 removed). Finally, since this study’s target population comprises OSS project contributors, we also inspected the answers to the question about years of experience in OSS to filter answers from participants with no experience (0 removed). After applying all the filters, we ended up with 242 valid responses.

\subsection{Data analysis} 
\label{section:method:data_analysis}

\begin{table*}[!htb]
\centering
\caption{Representative examples of answers to the open questions that were coded for each category}
\label{tab:codebook}
\begin{tabularx}{\textwidth}{L{0.08\textwidth}L{0.88\textwidth}}
\toprule
\textbf{Motivation}  & \textbf{Representative Examples}\\ \hline
 \midrule
Ideology & \textit{``I believed in the free software philosophy"} (P140);  \textit{``the ideology and community really appealed to me"} (P155); \textit{``open source was the most ethical"} (P82) 
\\\hline

Altruism & \textit{``{[}...{]} to support other FOSS contributors"} (P25); \textit{``To spread knowledge"} (P30); \textit{``to make the world better"} (P2)   \\\hline

Fun  & \textit{
``Because I enjoy coding"} (P69); \textit{``hobbyism"} (P49); \textit{``Because it's fun"} (P148). \\\hline

Kinship & \textit{``I liked the idea of collaborating with strangers on a project''} (P47) 
; \textit{``The community around the software library is very inclusive"} (P171)      \\\hline

Reputation  & \textit{``to improve my reputation"} (P21); \textit{``I saw F/OSS as a way to gain recognition"} (P90); \textit{``recognition"} (P76) \\\hline

Reciprocity    & 

\textit{``I benefit from it. It seemed right to to give back."} (P48); \textit{``I use the open source software, it's also great to contribute"} (P67); \textit{``I publish, and use, F/OSS projects daily, so that's just my duty {[}to give back{]}."} (P77)  \\\hline

Learning & \textit{``A good source for learning new things"} (P75); \textit{``learn from experienced developers from industry"} (P149); \textit{``Learn new things from different people"} (P182) \\\hline

Own-use   & \textit{``To help to improve software I used"} (P109);  \textit{``I wanted to solve problems that existed for me as a user"} (P132); \textit{``I end up tweaking the projects that I contribute for my own purposes"} (P149)  \\\hline

Career  & \textit{``Create a good public portfolio which can be good in hiring"} (P75); \textit{``I can put it on my CV"} (P240); \textit{``building a curriculum"} (P215)  \\\hline
Pay & \textit{``Because I get paid to continue to contribute"} (P132); \textit{``Motivated by employer"} (P195); \textit{``I received a stipend to begin"} (P137)   \\\hline
GSoC  & \textit{``I found the Google Summer of Code (GSoC) program and began contributing"} (P45); \textit{``A summer internship (GSoC)"} (P27)                                                        \\\hline
Coursework  & \textit{``In a college course \ldots students were encouraged to make contributions"} (P230); \textit{"a university class about F/OSS"} (P150) \\ 
\bottomrule
\end{tabularx}
\end{table*}

\subsubsection{Likert-scale questions} 
\label{section:method:data_analysis:closedquestions}
We used the Likert-scale items to compare our results to the previous surveys (RQ1). We ranked the questions based on the number of respondents who agreed to each motivation (checking ``Agree'' or ``Strongly Agree''). We then compared each previous papers' ranking with a corresponding ranking of our answers, built by excluding the items that did not match those from the previous study. We detail this analysis in Section \ref{section:results:rq1a}.

\subsubsection{Open questions}  
\label{section:method:data_analysis:openquestions}
To address RQ2 and RQ3, we analyzed the answers to the open questions about motivation to start and to continue contributing. We categorized the answers based on a card sorting approach~\cite{Spencer2009}. We used the categories from \citet{von2012carrots} (see Section~\ref{motivation_factors}) as a seed to the classification. We discarded four responses that left blank the answers about their ``motivation to start'' and ``motivation to keep contributing.'' However, we kept two cases in which respondents provided motivations only for one of them (the motivation to start in both cases). 

The whole process was conducted using continuous comparison~\cite{continuos_comparison} and discussion until reaching consensus. To include an outside view, we invited a researcher who had not participated in the method's initial steps to pair with one of the previous researchers. These two researchers jointly analyzed two sets of 20 answers to establish common ground, discussing the applied codes. Then, each researcher analyzed the remaining answers independently and discussed the disagreements until reaching consensus. Finally, a third researcher inspected the classification. Table~\ref{tab:codebook} presents representative examples for each category.

During this process, we decided to include two new categories that did not fit well in~\citet{von2012carrots}'s classification: \emph{``GSoC (Google Summer of Code)''} and \emph{``Coursework.''} The literature about joining OSS via GSoC and course assignments suggests that there are multiple motivations associated with these reasons to participate~\cite{SILVA2020110487,TheoryGSoC,8166691}.

\subsubsection{Segment analysis} 
\label{section:method:data_analysis:segmentedanalysis}

To analyze how the reported motivation differ according to individual characteristics, we segmented our sample based on \emph{experience in OSS} (experienced: fourth quartile, $\geq$ 15 years of experience vs. novices: first quartile, $\leq$ 3 years of experience), \emph{age} (older: $\geq$ median, 35 years old vs. younger: $<$ median, 35 years old), and \emph{role} (coder: `code developer' or `code reviewer' as one of the top-3 activities vs. non-coder: other activities).



\subsection{Replication package} 
A comprehensive replication package including our anonymized dataset, instruments, and scripts is stored in the Zenodo\footnote{https://doi.org/10.5281/zenodo.4453904} open data archive.

\section{Participant Demographics}
\label{section:demographics}

Our survey received 242 valid answers. In the following, we report the demographics of the respondents. No questions were mandatory, so not all categories sum to 242. The demographics are presented in Table~\ref{tab:demographics}.

\begin{table}[tbh]
\centering
\caption{Personal characteristics of the respondents (n=242)}
\label{tab:demographics}
\begin{tabular}{l|r|r}
\hline
\toprule
 \textbf{Demographics} & \textbf{\#}  & \textbf{\%}  \\
 \hline
\midrule
Gender: Man                                          & 196 & 82.7\% \\
Gender: Woman                                        & 18  & 7.6\%  \\
Gender: Non-binary                                   & 1   & 0.4\%  \\
Gender: Prefer to self describe                      & 1   & 0.4\%  \\
Gender: Prefer not to say                            & 21  & 8.9\%  \\
\toprule
Experience: $\leq$ 3 years in OSS                     & 63  & 27.9\% \\
Experience: $>$ 3 \& $<$ 15 years in OSS & 98  & 43.4\% \\
Experience: $\geq$ 15 years in OSS               & 65  & 28.8\% \\
\toprule
Age: 24 or less                                      & 42  & 18.7\% \\
Age: 25 to 34                                        & 71  & 31.6\% \\
Age: 35 to 44                                        & 73  & 32.4\% \\
Age: 45 to 54                                        & 30  & 13.3\% \\
Age: 55 to 64                                        & 8   & 3.6\%  \\
Age: Over 64                                         & 1   & 0.4\% \\
\toprule
Role: Coder                                          & 193 & 81.1\% \\
Role: Non-Coder                                      & 45  & 18.9\% \\
\toprule
Continent: North America                              & 70  & 31.8\% \\
Continent: South America                                & 25  & 11.4\% \\
Continent: Europe                                 & 100 & 45.5\% \\
Continent: Africa                                 & 2   & 0.9\%  \\
Continent: Asia                                   & 16  & 7.3\% \\
Continent: Australia                                & 7   & 3.2\%  \\
\toprule
Financial: Paid                              & 15  & 6.6\% \\
Financial: Unpaid                              & 140  & 61.4\% \\
Financial: Mostly paid                          & 26  & 11.4\% \\
Financial: Mostly unpaid                   & 27  & 11.8\% \\
Financial: Similar paid and unpaid      & 20  & 8.8\% \\

\bottomrule
\end{tabular}
\end{table}
\begin{figure*}[!b] 
  \includegraphics[width=\textwidth]{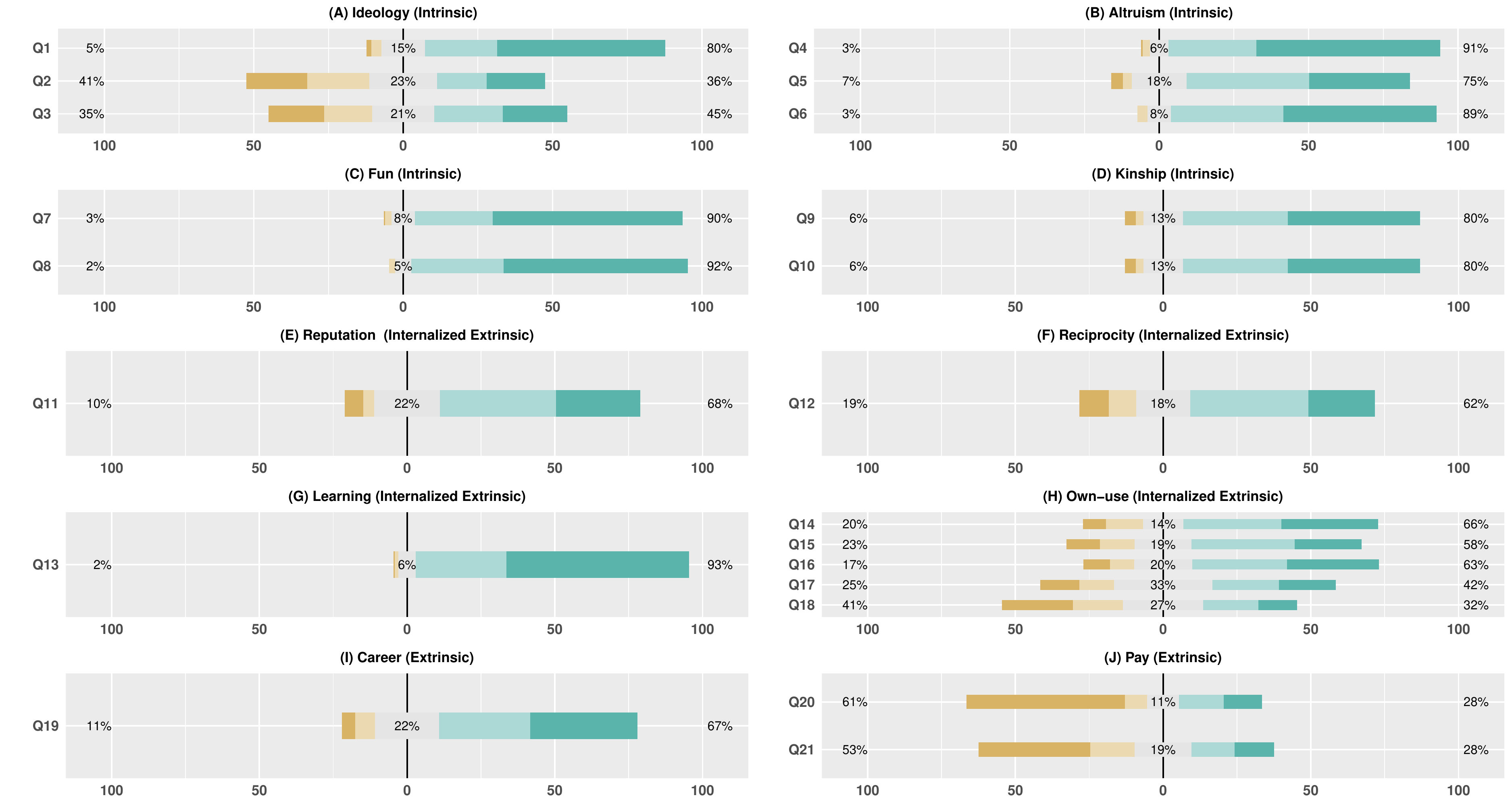}
  \caption{Responses to the 5-point Likert-scale items for motivation to contribute to OSS. Left hand (yellow) shows levels of disagreement, middle (grey) shows neutral, and right (green) shows levels of agreement.}
  \label{fig:likert_motivation}
\end{figure*}

We received answers from residents of five different continents with a broad age distribution. The majority are men (82\%) and coders (81\%) matching previously reported distributions of OSS contributors~\cite{GitHubOpenSourceSurvey2017, robles2013}. 


The population is also diverse in terms of the projects to which they contribute. The respondents reported contributing, for example, to the Linux kernel, KDE, Debian, Kubernetes, LibreOffice, Mozilla, PHP, Laravel, Drupal, Debian, TensorFlow, Apache projects, Firefox, Homebrew, Arduino, Eclipse, Joomla, Django, WordPress, JavaScript libraries, Python libraries, and R packages. The projects are diverse in programming languages, age, community size, organization, and governance model. 

Regarding the financial relationship with OSS, most respondents reported being unpaid, and only 26\% reported that they receive at least some money to work with OSS. On average, the respondents spend 10 hours per week (min=$<$1 hour, median=5, max=60) and have 9 years of experience with OSS (min=1 year, median=8, max=30).

\section{Results}
\label{section:results}

\subsection{RQ1a: What motivates contributors to OSS today?}
\label{section:results:rq1a}

Figure~\ref{fig:likert_motivation} shows the answers to the Likert-scale items about what motivates OSS contributors, grouped as per~\citet{von2012carrots}'s categories.


Figure~\ref{fig:likert_motivation} shows agreement among participants that intrinsic motivations, especially, \emph{Fun}, \emph{Altruism}, and \emph{Kinship}, are key motivations---on average 91\%, 85\%, and 80\% of the respondents agree (or strongly agree) that they contribute to OSS due to these motivations. This is reflected in P164's excitement of contributing to OSS: ``\textit{Discovered Linux and open source in general when I was a student in the 90s. Sending a patch across the ocean just seemed very exciting}" and P30's altruistic vision: ``\textit{To spread knowledge, which I think that contributes to make a society better}".




Internalized-extrinsic motivations---\emph{Learning} and \emph{Reciprocity}---are also important factors. An impressive 93\% of respondents (Figure~\ref{fig:likert_motivation}(G)) agreed that they contribute because OSS allows them to learn and improve their skills, as P75 explained: ``\textit{I continued contributing to OSS projects because it was a good source for learning new things.}"


Extrinsic factors like \emph{Career} and \emph{Pay}, paint a contrasting picture. While 67\% participants agree that OSS presents opportunity for professional growth (with only 11\% disagreement, Figure\ref{fig:likert_motivation}(I)), only 28\% mention payment as a motivation (61\% and 53\% disagreeing with Q20 and Q21, respectively, Figure~\ref{fig:likert_motivation}(J)).

\begin{table*}[!bt]
\centering
\scriptsize
\caption{Ranking comparisons: previous surveys vs. corresponding relative rankings of our results*}
\label{tab:literature_comparison*}
\resizebox{0.9\textwidth}{!}{%
\begin{tabular}{l|rl|!{\vrule width 1pt}r|r|!{\vrule width 1pt}r|r|!{\vrule width 1pt}r|r!{\vrule width 1pt}r}
\toprule
\textbf{Motivation}&\multicolumn{2}{l|}{\textbf{Question}} & \textbf{Hars} & \textbf{Ours H*} & \textbf{Lakhani} & \textbf{Ours L*} & \textbf{Ghosh} & \textbf{Ours G*} & \textbf{Ours} \\\hline
\midrule
\multirow{3}{*}{Ideology}& Q1&I believe that source code should be open   &  & & 4 & 3  & 4 & 3 & 6 \\
& Q2&I dislike proprietary software and want to defeat them               &   & & 9 & 10 &   &  & 18\\
& Q3&I want to limit the power of large software companies                &   & &   & & 8 & 9 & 16 \\\hline
\multirow{3}{*}{Altruism}& Q4&I want to share knowledge and skills                                 &   & &   & & 2 & 2 & 5 \\
& Q5&I want to improve the product of other developers                    &   & &   & & 3 & 5 & 3  \\
& Q6&I deeply enjoy helping others               & \cellcolor{gray!25} 7 & \cellcolor{gray!25} 2  &   & &   & & 9\\\hline
\multirow{2}{*}{Fun}& Q7&I have fun writing programs &   & &  &  &  & & 4\\
& Q8&I feel intellectually stimulated by writing code &   & & 1 & 2  &   & & 2 \\\hline
\multirow{2}{*}{Kinship}& Q9&I like to work with this(these) development team(s) & \cellcolor{gray!25} 6 &  \cellcolor{gray!25}3  & \cellcolor{gray!25} 7 & \cellcolor{gray!25} 3 &  & & 6 \\
& Q10&I want to participate in the F/OSS scene     & 2 & 3  &   & & 5 & 3 & 6 \\\hline

\multirow{1}{*}{Reputation}& Q11&I want to enhance my reputation  & 3 & 5  & \cellcolor{gray!25} 10  & \cellcolor{gray!25} 5  & \cellcolor{gray!25} 11   & \cellcolor{gray!25} 6  & 10 \\\hline
\multirow{1}{*}{Reciprocity}& Q12&I feel personal obligation because I use F/OSS                      &   & & 6 & 8  &   & & 14 \\\hline
\multirow{1}{*}{Learning}& Q13&I want to develop and improve my skills  & 1 & 1  & 2 & 1  & 1 & 1 & 1 \\\hline
\multirow{5}{*}{Own-Use}& Q14&I need the software for my work & \cellcolor{gray!25} 4 & \cellcolor{gray!25} 7  & \cellcolor{gray!25} 3 & \cellcolor{gray!25} 7  &   & & 12 \\
& Q15&I need the software for non-work purposes   &   & & \cellcolor{gray!25} 5 & \cellcolor{gray!25} 9  &   & & 15\\
& Q16&Problem could not be solved by proprietary software &   & &   & & 7 & 8 & 13 \\
& Q17&The projects that I contribute to would not make money  &   & &   & & 12   & 10 & 17 \\
& Q18&I need help in realizing a good idea for a software product   &   & &   & & 9 & 11 & 19\\\hline
\multirow{1}{*}{Career}& Q19&I want to improve my career opportunities   & 5 & 6  & 8 & 6  & 6 & 7 & 11 \\\hline
\multirow{2}{*}{Pay}& Q20&I am paid to contribute   &   & &   & & 10   & 12 & 20 \\
& Q21&Sell products and services related to F/OSS & 8 & 8  &   & &   &  & 20 \\
\bottomrule
\end{tabular}
}
\resizebox{0.9\textwidth}{!}{%
\begin{tabularx}{\linewidth}{@{}XXX@{}}

* Our ranking is recalculated for each comparison considering only the questions that are common to both surveys. We highlight the cells in which the motivation changed from one half to the other with a minimum difference of three positions. 
\end{tabularx}
}




\end{table*}

Finally, some of the original motivations for contributing to OSS---\emph{Ideology} \& \emph{Own-Use}---show mixed responses. Some aspects of ideology, such as opposing large companies and proprietary software, were not as popular as other motivations, which could be a result of large companies' recent embrace of OSS. On the other hand, the philosophy that source code should be open still remains strong (80\%, Figure~\ref{fig:likert_motivation}(A)). As P140 said ``\textit{I believe in the free software philosophy.}"

\emph{Own-Use}, is a mixed bag. ``Scratch your own itch" was a key rallying call in the early days of OSS and is still a motivating factor for some, as P140 said: ``\textit{I contribute for my own purposes.}" However, the sentiment has changed. The two own-use questions with the biggest difference in opinions are Q18 and Q16 in Figure~\ref{fig:likert_motivation}(H). Q18 relates to people seeking help from the community to realize their idea. While about 32\% find this to be the case, a larger majority 61\% show people joining existing communities. Interestingly, about 63\% find proprietary software to be limited (Q16), at least in providing the same level of features, as P63 mentions: ``\textit{Depending on proprietary software was severely limiting, and possibility, as with OSS we can fix our own bugs.}''




\MyBox{\textbf{Motivations to Contribute:} Intrinsic and internalized motivations explain what drives most of the contributors today. On the extrinsic end, \emph{Career} is relevant to many contributors, contrary to \emph{Pay} which only explains why less than one-third of the respondents contribute to OSS.}

\subsection{RQ1b: How has motivation to contribute shifted as OSS has matured?}

To analyze the shifts in motivation in relation to previous surveys, we do not directly compare the percentages since each survey measured or aggregated the data in slightly different ways. For example, \citet{lakhani2003hackers} report the percentage of individuals based on their top three motivations, whereas \citet{harsworking} report the percentages of the individuals who ranked high or very high on each motivation subcategories. To provide a basis for comparison, we pairwise compare relative rankings considering exclusively those questions that appear in both surveys, generating one relative ranking of our results for each comparison (Table Table~\ref{tab:literature_comparison*}). For each previous study, we compare the ranking extracted from the original paper to a relative ranking of our results, which excludes the questions not present in the previous study being compared. 

Results show social aspects have gained considerable importance. \emph{Deeply enjoying helping others} (Q6) was ranked in the $7^{th}$ position in Hars et al.'s survey and $2^{nd}$ in our relative ranking. \emph{Kinship} (Q9) ascended in the ranking from the $6^{th}$ to $3^{rd}$ place when we compare to Hars et al., and from $7^{th}$ to $3^{rd}$ compared to Lakhani et al.'s results. These questions were not asked in Ghosh et al.'s survey. 

There was also a shift in the relative importance of \emph{Reputation} (Q11). This motivation was ranked $10^{th}$ (last) in Lakhani et al. and $11^{th}$ (second to the last) in Ghosh et al., and moved up to the first half of our relative rankings: $5^{th}$ and $6^{th}$, respectively. However, the same trend does not hold when comparing to Hars et al.

There was a change in the opposite direction for the motivations related to ``scratch one's own itch" (\emph{Own-Use}). The question related to \emph{needing the software} was ranked $4^{th}$ in Hars et al. and $7^{th}$ in our relative ranking. In relation to Lakhani et al., needing the software to work was ranked $3^{rd}$ and for non-work purposes was ranked $4^{th}$; in our relative ranking these items dropped to $7^{th}$ and $9^{th}$, respectively. 

Several motivations are consistently top-ranked in all surveys. \emph{Learning} (Q13) was ranked first in our survey, Hars et al.'s, and Ghosh et al.'s rankings---whereas it was ranked $2^{nd}$ in Lakhani et al's. \emph{Intellectual stimulation} (Q8) ($2^{nd}$ in our overall ranking), \emph{sharing knowledge and skills} (Q4) ($3^{rd}$ in our overall ranking), \emph{participating in the OSS scene} (Q10) ($5^{th}$ in our overall ranking), and \emph{belief that source code should be open} (Q1) ($5^{th}$ in our overall ranking) were also top ranked in previous surveys. 

Alternatively, we can observe that financial reward (being paid to contribute and selling products and services) was bottom-ranked for all surveys.

\MyBox{\textbf{Shifting motivations through time:} 
Some motivations to contribute to OSS have stood the test of time: learning, fun, knowledge sharing, and a belief that source code should be open---all core tenets of OSS. Others have seen a marked difference. Social aspects (e.g., altruism, kinship, and reputation) have gone up in the ranking, whereas participating in OSS to ``scratch one's own itch'' has dropped. 
}

\subsection{RQ2: How does motivation to contribute to OSS shift as OSS contributors gain tenure?}
\label{section:results:rq2}

To analyze the shift in motivation from the perspective of the contributors, we asked two open-ended questions about what motivated them to \textit{start} contributing and why they \textit{continued} to contribute. We qualitatively analyzed the answers using \citet{von2012carrots}'s categories as seeds, following the procedures described in Section \ref{section:method:data_analysis:openquestions}. 

\begin{figure}[!ht]
\scriptsize
\centering
     \includegraphics[trim=0 2.5cm 0 2.5cm, clip=true, width=0.49\textwidth,height=8cm]{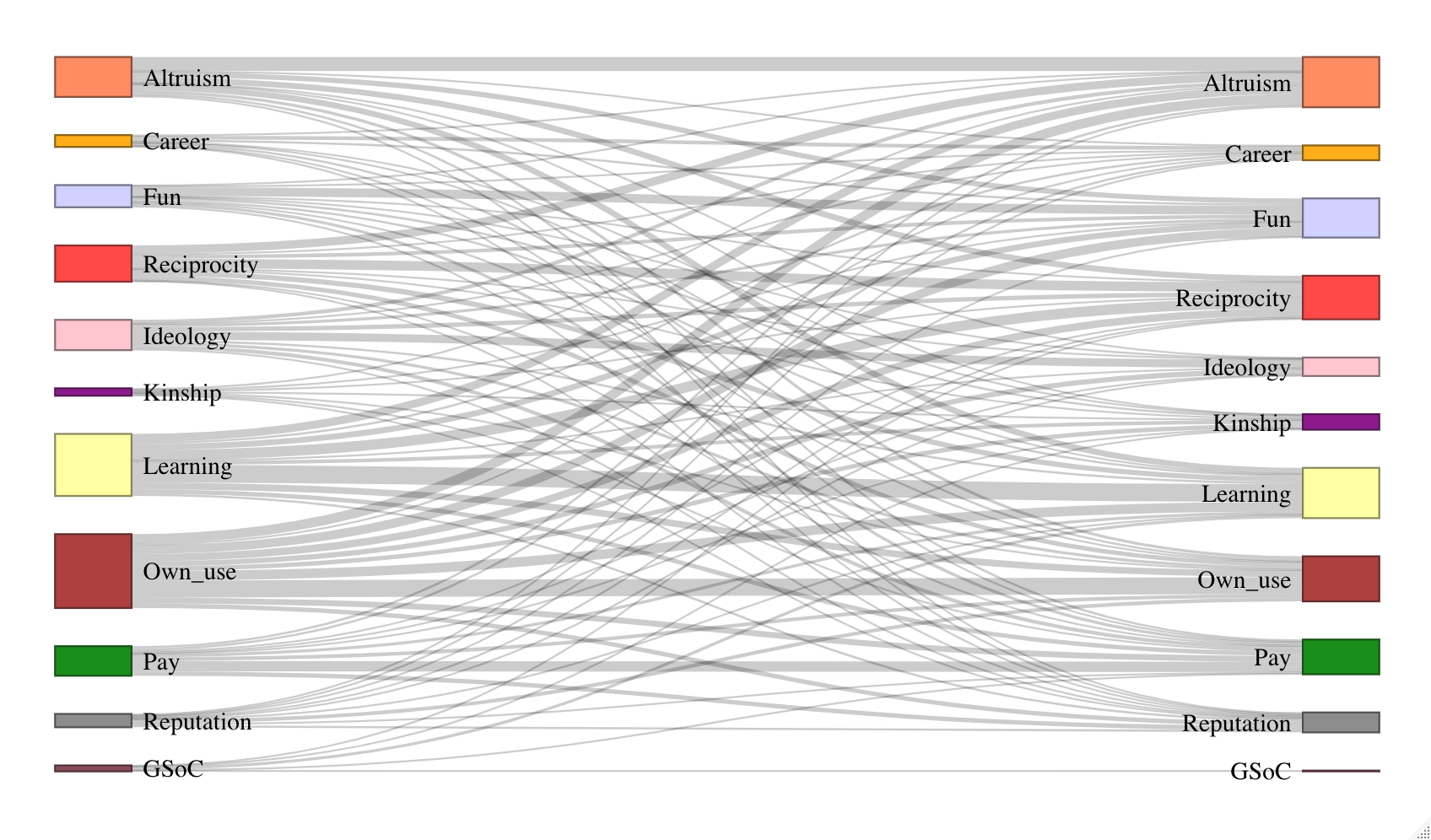}
    \caption{Motivation migration flow. The size of the boxes on the left represents the number of contributors with the motivation to start, and on the right to continue, contributing to OSS. The width of the connections is proportional to the number of contributors who shifted from one motivation to the other.}
    \label{fig:motivation_migration}
\end{figure}

Figure~\ref{fig:motivation_migration} shows how the motivations of our respondents shifted and Table~\ref{tab:attributes} shows how often each motivation was cited as a reason to start and to continue contributing. \emph{Fun} was mentioned by more participants (18.9\%) as a reason to continue than to start (9.2\%). The frequency of mentions to \emph{Altruism} also increased, from 18.0\% to 26.7\%. \emph{Reputation}, \emph{Kinship}, and \emph{Reciprocity} also had noticeable increases. Finally, we noticed that \emph{Pay} was a more common reason to continue than to join, shifting from 12.4\% to 16.6\%. 

On the opposite side, \emph{Own-Use} experienced the largest drop (decreasing from 29.0\% to 21.7\%). \emph{Ideology} also lost positions, decreasing from 13\% to 9\%. \emph{Summer of code programs} and \emph{Coursework} also were more common reasons to start than to continue contributing. 

In terms of individual migrations, the contributors who joined motivated by \emph{Own-Use} commonly migrated to intrinsic and internalized extrinsic motivations---\emph{Altruism} (20.3\%), \emph{Learning} (20.3\%), \emph{Fun} (18.8\%), and \emph{Reciprocity} (15.6\%). 
Similarly, 43\% of those who started because of \emph{Pay} (extrinsic) migrated to internalized motivations, and 61\% of those who started because of \emph{Career} (extrinsic) migrated to intrinsic or internalized motivations. The same happened for \emph{Summer of Code Programs} and \emph{Coursework}. Finally, those who started in OSS because of \emph{Ideology} migrated to \emph{Reciprocity} (20\%), \emph{Altruism} (16.7\%), and \emph{Pay} (16.7\%). The full list of migration flows is available in the supplementary material.

\MyBox{\textbf{Shifting motivations in self-journeys:} Contributing to OSS often transforms extrinsic motivations to intrinsic ones. Whereas ideology, own-use or education-related programs can be an impetus to join OSS, individuals continue for intrinsic reasons (fun, altruism, reputation, and kinship). 
}

\begin{table}[!bt]
\centering
\begin{threeparttable}
\caption{Number of respondents who reported each motivation to start and to continue contributing *}
\label{tab:attributes}
\begin{tabular}{lrrr}
\toprule
\textbf{{Motivation}}&\textbf{\# start} & \textbf{\# continue} & \textbf{Difference}\\\hline
\midrule
Fun & 20 (9.2\%) & 41 (18.9\%) &105\% $\Uparrow$ \\
Altruism & 39 (18.0\%) & 58 (26.7\%) &49\% $\Uparrow$ \\
Reputation & 9 (4.1\%) & 23 (10.6\%) &156\% $\Uparrow$ \\
Kinship & 7 (3.2\%) & 19 (8.8\%) &171\% $\Uparrow$ \\
Reciprocity & 28 (12.9\%) & 39 (18.0\%) &39\% $\Uparrow$ \\
Pay & 27 (12.4\%) & 36 (16.6\%) &33\% $\Uparrow$ \\
Career & 9 (4.1\%) & 15 (6.9\%) &67\% $\Uparrow$ \\
Learning & 45 (20.7\%) & 49 (22.6\%) &9\% $\Uparrow$\\\hdashline
Coursework & 2 (0.9\%) & 0 (0.0\%) &100\% $\Downarrow$\\
GSoC & 5 (2.3\%) & 1 (0.5\%) &80\% $\Downarrow$\\
Ideology & 28 (12.9\%) & 19 (8.8\%) &32\% $\Downarrow$\\
Own-use & 63 (29.0\%) & 47 (21.7\%) &25\% $\Downarrow$\\
\bottomrule
Total & 281 & 347 & -- \\
\bottomrule 
\end{tabular}
\begin{tablenotes}
    \scriptsize
      \item *The motivations were coded from the open questions. A participant could list more than one motivation to start or continue contributing. The table is sorted by the difference between the relative frequencies (percentages) of the motivations to start and continue (not to be confused with the percentage value of the difference of the absolute numbers (last column)).
    \end{tablenotes}
\end{threeparttable}

\end{table}
\subsection{RQ3a: How does motivation to contribute differ for diverse characteristics?}

Table~\ref{tab:odds} presents the odds ratio for different subgroups of respondents reporting each motivation factor as a reason to continue contributing to OSS (survey open question). Experienced developers have higher odds to report \emph{Altruism} (5.6x), \emph{Pay} (5.2x), and \emph{Ideology} (4.6x) than novices. On the other hand, novices have greater odds to report \emph{Career} (10x), \emph{Learning} (5.5x), and \emph{Fun} (2.5x).

The odds that contributors who are 35 years or older report \emph{Pay} and \emph{Altruism} are 4.1x and 2.1x higher than for younger contributors. Respondents younger than 35 have higher odds to report \emph{Learning} (3.3x) as a reason to contribute.

The odds of \emph{Fun} motivating coders were 4x higher than non-coders. Still, non-coders have higher odds of mentioning \emph{Ideology} (2.5x) as a motivator, probably because of the type of activity (e.g., ``advocates and evangelists''). 

\begin{table}[!tb]
\centering
\begin{threeparttable}

\caption{Odds ratios per personal characteristic}
\label{tab:odds}
\begin{tabular}{l|l|l|l|l}
\toprule
& \textbf{\begin{tabular}[x]{@{}l@{}}Experienced vs.\\ Novice\end{tabular}} & \textbf{
\begin{tabular}[x]{@{}l@{}}Older vs.\\ Younger\end{tabular}}  & 
\textbf{\begin{tabular}[x]{@{}l@{}}Men vs.\\ Women\end{tabular}}
& \textbf{
\begin{tabular}[x]{@{}l@{}}Coder vs.\\ Non-coder\end{tabular}} \\
\hline  
\midrule
Ideology         & \cellcolor{gray!25}4.6**       & 1.1   & 1.3      & 0.4  \\
Altruism         &\cellcolor{gray!25}5.6**       & \cellcolor{gray!25}2.1**   & 1.1   & 0.5*   \\
Fun        & \cellcolor{gray!25}0.4**       & 0.5   & 4.3      & \cellcolor{gray!25}4.0**   \\
Kinship          & 2.1       & 0.8   & 0.6      & 1.0   \\
Reputation       & 0.5       & 0.9   & 1.5      & 0.9   \\
Reciprocity & 1.0       & 0.7   & \cellcolor{gray!25}0.3**      & 0.7   \\
Learning         & \cellcolor{gray!25}0.2**       & \cellcolor{gray!25}0.3**   & 1.2      & 1.1   \\
Own-Use          & 0.9       & 0.6   & 4.3      & 1.6  \\
Career           &\cellcolor{gray!25}0.1**       & 0.6   & 1.4      & 1.8   \\
Pay              & \cellcolor{gray!25}5.2**       & \cellcolor{gray!25}4.1**   & 1.2      & 1.1   \\
\bottomrule
\multicolumn{4}{l}{\footnotesize Significance codes: * $p<0.10$, ** $p<0.05$.}\\

\end{tabular}

\begin{tablenotes}
    \scriptsize
      \item Note: Odds ratio greater than 1 means that the first segment has greater chances of reporting the motivation than the second. Ratio less than 1 means the opposite. The motivations were coded from the survey open questions.
\end{tablenotes}
\end{threeparttable}
\end{table}

\MyBox{\textbf{Individual characteristics affect motivations:}
Social and philosophical factors (such as \emph{Ideology} and \emph{Altruism}) as well as  \emph{Pay} are reported more often by experienced developers, whereas professional factors (e.g., \emph{Career}), \emph{Learning}, and \emph{Fun} are more common among novices. 
}

\subsection{RQ3b: How do shifts in motivation to contribute differ for diverse characteristics?}
\label{section:results:rq3}

To understand how shifts in motivation relate to individual characteristics, we segmented the data from Table~\ref{tab:attributes} (reasons for starting to contribute and reasons for continuing to contribute as reported in the survey open questions). Table~\ref{tab:demographics_before_after} shows the result of this segment analysis.

\begin{table*}[htb]
\centering
\scriptsize
\caption{Number of individuals who Started$\rightarrow$Continued because of each motivation by personal characteristics *.}
\label{tab:demographics_before_after}
\resizebox{0.97\textwidth}{!}{%

\begin{tabular}{l|r|r|r|r|r|r|r|r|r|r}
\hline
\toprule
\textbf{Demographics / Motivations}    
& \textbf{Ideology} & \textbf{Altruism} & \textbf{Fun} & \textbf{Kinship} & \textbf{Reputation} & \textbf{Reciprocity} & \textbf{Learning} & \textbf{Own-Use} & \textbf{Career} & \textbf{Pay} \\ \hline
 \midrule
Gender:Man (n=196)     & 
\cellcolor{red!25} 22$\rightarrow$14  & \cellcolor{blue!25} 34$\rightarrow$48 & \cellcolor{blue!25} 13$\rightarrow$40 & \cellcolor{blue!25} 6$\rightarrow$16 & \cellcolor{blue!25} 7$\rightarrow$19  & \cellcolor{blue!25} 23$\rightarrow$29 & 33$\rightarrow$40 & 50$\rightarrow$43 & \cellcolor{blue!25} 9$\rightarrow$16 & 24$\rightarrow$28 \\

Gender:Woman (n=18)     & 
2$\rightarrow$1  & \cellcolor{blue!25} 0$\rightarrow$4    & 0$\rightarrow$1   & 0$\rightarrow$2  & 0$\rightarrow$1 & \cellcolor{blue!25} 1$\rightarrow$7 & 2$\rightarrow$3 & \cellcolor{red!25} 6$\rightarrow$1 & 0$\rightarrow$2 & 2$\rightarrow$2 \\
\toprule

OSS Exp:High($\geq$15) (n=65)  & 
10$\rightarrow$8 & \cellcolor{blue!25} 10$\rightarrow$22 & 6$\rightarrow$9 &  3$\rightarrow$4 & 4$\rightarrow$5 & 8$\rightarrow$11 & 7$\rightarrow$5 & \cellcolor{red!25} 17$\rightarrow$11 & 2$\rightarrow$1  & \cellcolor{blue!25} 5$\rightarrow$16 \\

OSS Exp:Medium($>$3\&$<$15) (n=98) & 
\cellcolor{red!25} 13$\rightarrow$9 & 22$\rightarrow$27 & \cellcolor{blue!25} 2$\rightarrow$16 & \cellcolor{blue!25} 2$\rightarrow$6 & \cellcolor{blue!25} 1$\rightarrow$9 & \cellcolor{blue!25} 12$\rightarrow$16  & 15$\rightarrow$15 & \cellcolor{red!25} 25$\rightarrow$21 & 3$\rightarrow$5 & 11$\rightarrow$13\\

OSS Exp:Low($\leq$3) (n=63)   & 
2$\rightarrow$2 & 6$\rightarrow$5  & \cellcolor{blue!25} 6$\rightarrow$19 &  \cellcolor{blue!25} 2$\rightarrow$10 & \cellcolor{blue!25} 2$\rightarrow$6 & \cellcolor{blue!25} 6$\rightarrow$11  & \cellcolor{blue!25} 15$\rightarrow$24 & \cellcolor{red!25} 18$\rightarrow$12 & \cellcolor{blue!25} 4$\rightarrow$12  & 7$\rightarrow$4\\
\toprule

Age:($\geq$35 yrs) (n=112)   & 
13$\rightarrow$10 & \cellcolor{blue!25} 19$\rightarrow$35 & \cellcolor{blue!25} 9$\rightarrow$16 & 5$\rightarrow$8 & \cellcolor{blue!25} 5$\rightarrow$11 & 15$\rightarrow$16 & 14$\rightarrow$12 & \cellcolor{red!25} 24$\rightarrow$18  & 3$\rightarrow$6 & \cellcolor{blue!25} 17$\rightarrow$24\\

Age:($<$35 yrs) (n=113)   & 
12$\rightarrow$9 & 17$\rightarrow$17  & \cellcolor{blue!25} 6$\rightarrow$27 & \cellcolor{blue!25} 2$\rightarrow$12 & \cellcolor{blue!25} 2$\rightarrow$9 & 1\cellcolor{blue!25} 1$\rightarrow$21 & \cellcolor{blue!25} 23$\rightarrow$32 & \cellcolor{red!25} 35$\rightarrow$26 & \cellcolor{blue!25} 6$\rightarrow$12 & 9$\rightarrow$9\\
\toprule

Role:Coder (n=193)     & 
\cellcolor{red!25} 20$\rightarrow$13 & \cellcolor{blue!25} 30$\rightarrow$41 & \cellcolor{blue!25} 15$\rightarrow$43 & \cellcolor{blue!25} 4$\rightarrow$17 & \cellcolor{blue!25} 6$\rightarrow$19 & \cellcolor{blue!25} 22$\rightarrow$30 & \cellcolor{blue!25} 32$\rightarrow$40 & \cellcolor{red!25} 56$\rightarrow$41  & \cellcolor{blue!25} 9$\rightarrow$17 & \cellcolor{blue!25} 23$\rightarrow$30\\

Role:Non-Coder (n=45)     & 
7$\rightarrow$7 & \cellcolor{blue!25} 8$\rightarrow$15 & 3$\rightarrow$3 & 3$\rightarrow$4 & 2$\rightarrow$4  & 6$\rightarrow$9 & 8$\rightarrow$10 & 7$\rightarrow$6 & 0$\rightarrow$2  & 5$\rightarrow$6\\
\toprule

Residence:North America (n=70)    & 
9$\rightarrow$6 & \cellcolor{blue!25} 13$\rightarrow$20 & \cellcolor{blue!25} 2$\rightarrow$10 & 3$\rightarrow$6 & 3$\rightarrow$5 & \cellcolor{blue!25} 6$\rightarrow$12 & 11$\rightarrow$8 & \cellcolor{red!25} 17$\rightarrow$13 & 2$\rightarrow$5  & \cellcolor{blue!25} 11$\rightarrow$15\\

Residence:South America (n=25)    & 
3$\rightarrow$4 & 4$\rightarrow$5  & 1$\rightarrow$3 & 0$\rightarrow$1 & 1$\rightarrow$3 & 4$\rightarrow$5  & \cellcolor{blue!25} 3$\rightarrow$7 & 6$\rightarrow$5 & 1$\rightarrow$2 & 3$\rightarrow$2 \\

Residence:Europe (n=100)     & 
9$\rightarrow$7 & 18$\rightarrow$22 & \cellcolor{blue!25} 10$\rightarrow$21 & \cellcolor{blue!25} 3$\rightarrow$9 & \cellcolor{blue!25} 3$\rightarrow$8 & 14$\rightarrow$17 & 16$\rightarrow$17 & \cellcolor{red!25} 32$\rightarrow$23 & 4$\rightarrow$7  & \cellcolor{blue!25} 9$\rightarrow$15\\

\bottomrule
\end{tabular}
}
\resizebox{0.97\textwidth}{!}{%
\begin{tabularx}{\linewidth}{@{}XXX@{}}
\\** In each cell, the number on the left represents the number of responses mentioning the motivation as a reason to start while the number on the right, as a reason to continue. We excluded from the table the \emph{GSoC} and \emph{Coursework} categories due to the limited number of respondents. We also excluded the continents and genders for which we had few responses. We highlight in \colorbox{blue!25}{blue}/\colorbox{red!25}{red} those cases in which we observed an increase/decrease of at least 25\% when comparing the second number (motivation to continue) in relation to the first (motivation to start) and at least 3 individuals.
\end{tabularx}
}

\end{table*}




There was a considerable increase (120\%) in \emph{Altruism} for experienced respondents ($\geq$15 years of experience in OSS), and a steady number (if not a slight decrease) for novices ($\leq$ 3 years of experience). The low number of novices who started because of \emph{Altruism} is also noticeable. The same trends are observed when analyzing the respondents' age.
 
\emph{Pay} was also a common reason to continue for experienced respondents. On the other hand, just a few young respondents joined OSS because of \emph{Career}, but many of them shifted towards this motivation---the number doubled when analyzing \emph{Career} as a reason to continue. \emph{Fun} also retains more young and novice contributors than older and experienced ones. 

We also segmented the data according to the contributors' role. More coders shifted their motivations than non-coders. The only case where we observed a higher increase for non-coders (88\%) was for \emph{Altruism}. An interesting finding here is that none of the non-coders mentioned \emph{Career} as a reason to join OSS, but two continued because of this motivation. On the other hand, the decreases in \emph{Ideology} and \emph{Own-Use} were more common for coders.


Finally, we analyzed the differences by continents. Overall, the shift in motivation was similar for North Americans, South Americans, and Europeans (we did not analyze the other continents due to their low number of responses). We found more noticeable differences for the shifts to intrinsic motives: \emph{Altruism}, \emph{Fun}, and \emph{Reciprocity}, which were bigger for North Americans (54\%, 400\%, 100\%) than for Europeans (22\%, 110\%, 21\%) and South Americans (25\%, 200\%, 25\%). 

\MyBox{\textbf{Shifting motivations in demographics:} 
Shifts in motivation as individuals continue to contribute are similar across demographics.
One notable difference is in experience; experienced contributors continue because of \emph{Pay} or \emph{Altruism}; but novices pursue for \emph{Career}. 
}


\section{Discussion}
We used~\citet{von2012carrots}'s work as our theoretical framework to analyze our data. 
\citet{von2012carrots} reviewed the literature published until 2009 (40 primary studies). While analyzing the responses to our open questions (more than 200 answers per question), the ten categories proposed in their work were adequate for coding the reported motivations. The few exceptions were Google Summer of Code and Coursework, with a few mentions each. Summer of code programs intend to attract contributors to OSS, and joining such programs involves several motivations, such as money, reputation, learning, and intellectual stimulation~\cite{TheoryGSoC,8166691}. Contribution to OSS has also been investigated as a way to foster learning and attract new contributors~\cite{githubemerge,8166691}. Next, we discuss some of the key findings in our survey results.

\textbf{Scratching one's own itch is not as big.} We observed a relative decrease of \emph{Own-Use}, both when comparing to prior surveys and analyzing the contributors' shifts, regardless of their individual characteristic. Those who joined OSS for \emph{Own-Use}-related reasons often shifted to \emph{Altruism}, \emph{Learning}, \emph{Fun}, and \emph{Reciprocity}, i.e., intrinsic/internalized motivations. Therefore, we could observe that intrinsic motivations have gained importance as OSS matured. 

\textbf{Learning is still a strong motivator.} \emph{Learning} continues to be a top motivator---it was top-ranked in ours and in prior surveys. In fact, it was ranked as the second most mentioned motivation to join and the first to continue. This shift towards learning was observed for all individual characteristics except for experienced contributors. It is reasonable to expect that highly-experienced contributors ($\geq$ 15 years) are core members who benefit less in terms of learning, but continue to contribute because they are paid or can help others.

\textbf{Payment and money are still not pervasive.} We expected financial reward to be a common motivation considering the increased involvement of industry~\cite{robles2019twenty}. However, it was bottom-ranked in our results. Nevertheless, as one could expect, the shift to this motivation was more common for experienced contributors. 

\textbf{There is a shift from extrinsic motivations to intrinsic ones.} Our results align with self-determination theory~\cite{deci2004handbook}, which helps to understand the dominant role intrinsic motivation plays. \citet{schmidt2017work} argue that individuals have a natural tendency to strive for a balance of externally-rewarded labor and intrinsically-rewarding leisure. Research shows that after an initial extrinsically motivated and challenging endeavor, motivation may shift towards a rewarding and intrinsically motivated task~\cite{kool2014labor}. \citet{Helen} show that motivation and job characteristics (e.g., pay and recognition) directly influence happiness and job satisfaction, which affects the intention to stay. According to these authors, although work motivation and job satisfaction are distinct, they are closely connected via a feedback loop, through a self-regulation process \cite{motivatingEmployees}. Our results reflect this movement, with OSS contributors shifting their motivation towards intrinsic or internalized factors. 



\textbf{Experienced members' motivations shifted towards altruism and social interaction.} Our results are also in line with Socioemotional Selectivity Theory (SST), which proposes that with age, the relative importance of goals shifts as a function of future time perspective~\cite{carstensen2006influence, carstensen1999taking}. According to this theory, younger adults focus more on future-oriented and horizon-expanding goals, like acquiring knowledge, whereas older adults focus more on present-oriented and emotionally meaningful goals such as maintaining high quality social bonds~\cite{steltenpohl2019me}. Indeed, our older and experienced respondents often shifted their motivation to \emph{Altruism}-related motives. On the other hand, younger adults and novices gravitate towards self-related motivations. 

Therefore, improving social aspects of the tools and projects can help retain developers as they age and become more experienced. Indeed, our study found that social aspects, such as helping others and working in teams, gained considerable relative importance compared to surveys from the early 2000s. Our results are also consistent with the findings from \citet{HelenESEM}, who reported ‘people’ as a common motivator. The emergence\footnote{https://octoverse.github.com/} of social coding platforms~\cite{tsay2} may explain this shift. These platforms offer a variety of social features and helped change the culture of OSS from a hacker-oriented group to a collaborative community, with lower barriers to entry and better support for newcomers~\cite{lower_barrier,mendez2018open}. 

\textbf{Novices want to promote their career with OSS.} Our results also show that young and novice contributors often shift their motivation towards \emph{Reputation} and other career-related motivations. This may relate to the new landscape of OSS~\cite{robles2019twenty, steinmacher2017free}, in which companies are key players. Moreover, novices may use their OSS contribution history as a portfolio, and potential employers are increasingly referring to online contributions when making hiring decisions~\cite{7111876}. 

\subsection{Implications and Future Work}

\textbf{To OSS Projects and Tool Builders:} 
Experienced contributors often shift their motivation towards \emph{Altruism}, valuing present-oriented and emotionally meaningful goals, such as maintaining high-quality social bonds. OSS projects willing to retain these experienced contributors, who tend to be core members or maintainers, could invest in strategies and tools showing how their work benefits the community and society, such as who uses the developed features. 

In response to the increased rank of ``I deeply enjoy helping others'' (Q6 in Table~\ref{tab:literature_comparison*}), GitHub and other platforms could offer social features to pair those needing help with those willing to help, highlight when a contributor helped someone, and make it easier to show appreciation to others (similar to stars given to projects). Projects and tool builders should also continue to facilitate collaboration and awareness \cite{steinmacher2010awareness}, in response to the increased relevance of ``I like to work with this (these) development team(s)'' (Q9 in Table~\ref{tab:literature_comparison*}). 

To attract and retain novices, who might become the future workforce, projects could invest in promoting \emph{Career}, \emph{Fun}, \emph{Kinship}, and \emph{Learning}, which are particularly relevant for novices and young contributors (see RQ3a/b). Mentors can  leverage these motivation factors to design or adapt specific strategies \cite{steinmacher2012recommending,steinmacher2018let}. Tools like Visual Resume~\cite{sarma2016hiring} can help novices promote their career by building their portfolios based on their contributions. Non-coders also play important roles in OSS communities~\cite{trinkenreich2020pathways} and their motivation also shifted towards \emph{Career} after joining. Their contributions should also be recognized in these enhanced portfolios. Current developer profile aggregators based on OSS data still focus on programming alone, e.g., conceptualizing skills in terms of a list of programming languages~\cite{singer2013mutual}. Adding non-coder contributions to such aggregators would be an important step towards recognizing a more diverse set of contributions.

Although diversity positively impacts OSS projects, most of the contemporary OSS projects lack diversity~\cite{bosu2019diversity}. It is important for OSS projects to understand and support contributors with different motivations. 

OSS communities may also invest in extrinsic motivation initiatives to attract contributors. This will pay off, since we observed that contributors will often progressively shift toward intrinsic motivations. In that sense, Coursework and summers of code programs can be interesting doorways for OSS; therefore, communities should put effort into welcoming students and applying to programs like GSoC.

\textbf{To Educators:} 
Employers increasingly refer to online contributions when making hiring decisions. OSS offers great potential to train the next generation of professionals~\cite{sarma2016hiring}. Young and novice contributors often shift towards professionally important motivations, such as \emph{Reputation} and \emph{Career}. Educators should offer coursework related to contributing to OSS and discuss contribution to OSS as part of software engineering training.

\textbf{To Researchers:} 
Motivation shifts have been poorly investigated in the OSS literature. Previous research has shown that motivation affects behavior, task effort, retention, and participation level. Further research is necessary to understand how the motivation shifts that we identified impact these constructs as well as disengagement and career trajectories. Researchers can replicate our study and further explore our data, which has been made available in our replication package. Finally, future studies can investigate the reasons behind the shifts and trends we observed. 



\subsection{Threats to Validity}

\textbf{Sampling bias.} In our case, random sampling is not viable, since there is no single list of all OSS contributors. We combined multiple strategies to reach a broad and diverse sample, as explained in Section~\ref{section:method:dist}. As described in Section~\ref{section:demographics}, we achieved a diverse population in terms of countries, projects, contribution roles, etc. Although the distribution of countries resembles the distribution of OSS contributions, there is a risk of a country bias. In terms of countries, USA (58) was dominant in North America (70), while Germany (23), UK (19), and Spain (18) are the most represented in Europe (100). We also have a small low number of women and non-binary respondents, which mirrors our population's characteristic lack of diversity~\cite{bosu2019diversity}. Furthermore, we acknowledge that our sample may be biased in unknown ways, and our results are only valid for our respondents.  

\textbf{Sample size}. We answered RQ3 by segmenting the dataset to analyze motivations and their shifts. In doing so, the dataset gets divided into smaller groups which could pose a challenge for statistical analysis. However, the group sizes we have (at least 15 responses per group) are sufficient for the conservative Fisher’s Exact Test \cite{cochran1952}, the underlying significance test used in the odds ratio analysis (RQ3a). In the descriptive analysis (RQ3b), we highlight only the most expressive shifts (those changing at least 25\%). 

\textbf{Response biases}. As in any survey method, our work can have recall bias---respondents answer only what they recall and not necessarily what was most important to them in the past. Recency and salience can also affect the respondents' answers. We aimed to reduce priming respondents with specific motivation factors by first presenting them with open questions, which allowed us to collect spontaneous answers.

\textbf{Survivability bias}. We focused our study on current OSS contributors. The motivations of those who tried but abandoned contributing may differ.

\textbf{Self-selection bias.} Participants decided whether they wanted to participate in the survey, and this may have influenced our results. Although most international OSS projects adopt English as their primary language, the language of the instrument may have influenced the willingness to participate of non-native speakers. Future studies might translate our survey and investigate regional differences. 

\textbf{Inappropriate participation.} As described in Section~\ref{section:method:filtering}, we employed several filtering and inspecting strategies to reduce the possibility of redundant participation and fake data; however, it is not possible to claim that our data is completely free of this threat. 

\textbf{Construct validity.} To enhance construct validity, we based our survey on previous instruments. However, these instruments were not formally validated and may inadequately measure a given motivation. To mitigate this threat, we employed pilot studies to test and collect feedback about our instrument.

\textbf{Subjectivity.} We employed qualitative procedures to classify the answers to the open questions and map questions from the previous surveys. These procedures are subject to subjectivity/interpretation bias. To mitigate this threat, we employed multiple researchers with diverse backgrounds, constant comparison, and negotiated agreement. All the researchers have extensive experience both in qualitative methods and OSS.


\section{Conclusion}


The ``sands of motivations'' shift over time and differ across demographics. To answer our research questions, we employed one of the most comprehensive surveys about motivation in OSS, encompassing questions from three seminal studies. Results show that social aspects, such as helping others, teamwork, and reputation have gained importance, while some intrinsic or internalized motivations are still prevalent, such as \emph{Learning}, \emph{Fun}, and \emph{Altruism}. Interestingly, OSS contributors often join because of extrinsic factors, but continue because of intrinsic factors. 

OSS projects can leverage our results to devise and review strategies to support each person to achieve their goals, resulting in more people engaging in OSS communities. Our results also shed light on the association between demographics and motivations, which, as fostering diversity is highly relevant to OSS communities, is important to consider in further research. 

\section*{Acknowledgments}
This work is partially supported by the National Science Foundation under Grant numbers 1815486, 1815503, 1900903, and 1901031, CNPq grant \#313067/2020-1, the Government of Spain through project ``BugBirth'' (RTI2018-101963-B-100), and the Australian Research Council’s Discovery Early Career Researcher Award (DECRA) funding scheme (DE180100153).
We also thank the Open Source contributors who spent their time answering our survey. We expect that the results benefit the OSS communities to help their growth and sustainability. 





\bibliographystyle{IEEEtranN}
\bibliography{reference}

\end{document}